\documentclass[12pt]{iopart}
\pdfoutput=1
\usepackage{graphicx,color}
\usepackage{cite}
\newcommand{\paddyspeaks}[1]{{\color{black} #1}}
\newcommand{\wkspeaks}[1]{{\color{black} #1}}

\begin{document}

\title[Locally favoured structures and dynamic length scales]{Locally favoured structures and dynamic length scales in a simple glass-former}

\author{C. Patrick Royall}
\address{H.H. Wills Physics Laboratory, Tyndall Avenue, Bristol, BS8 1TL, UK}
\address{School of Chemistry, University of Bristol, Cantock's Close, Bristol, BS8 1TS, UK}
\address{Centre for Nanoscience and Quantum Information, Tyndall Avenue, Bristol, BS8 1FD, UK}
\address{Department of Chemical Engineering, Kyoto University, Kyoto 615-8510, Japan}
\ead{paddy.royall@bristol.ac.uk}

\author{Walter Kob}
\address{Laboratoire Charles Coulomb,  UMR 5521, University of Montpellier and CNRS, 34095 Montpellier, France}
\ead{walter.kob@umontpellier.fr}

\vspace{10pt}
\begin{indented}
\item[]August 2016
\end{indented}

\begin{abstract}
We investigate the static and dynamic properties of a weakly polydisperse hard sphere system in the deeply supercooled state, \emph{i.e.}~at densities higher than that corresponding to the mode-coupling transition. The structural analysis reveals the emergence of icosahedral locally favoured structures, previously only found in trace quantities. We present a new approach to probe the shape of dynamically heterogeneous regions, which is insensitive to particle packing effects that can hamper such analysis. Our results indicate that the shape of the dynamically heterogeneous regions changes only weakly and moreover hint that the often-used four-point correlation length may exhibit a growth in excess of that which our method identifies. The growth of the size of the dynamically heterogeneous regions appears instead to be in line with the one of structural and dynamic \emph{propensity} correlations.
\end{abstract}

\section{Introduction}

Among the challenges of the glass transition is to understand how solidity, in the sense that no flow occurs on any relevant timescale, emerges with little apparent change in structure as seen in two point correlation functions. In fact, whether the glass transition has a thermodynamic (implying structural) or dynamical origin remains a matter of
debate~\cite{cavagna2009,berthier2011,binder}.  It has been proposed that upon cooling, icosahedral-like arrangements of atoms might form in the glassformer~\cite{frank1952} and that the dynamical arrest may be related to a (geometrically frustrated) transition to a phase of such icosahedra~\cite{tarjus2005}. To test these ideas studies have been carried out to identify geometric motifs such as icosahedra and related \emph{locally favored structures} (LFS) by means of computer simulation
\cite{steinhardt1983,jonsson1988,dzugutov2002,coslovich2007,eckmann2008,lerner2009,sausset2010,tanaka2010,malins2013jcp,charbonneau2012,charbonneau2013jcp,hocky2014,royall2015,dunleavy2015,pinney2015} and particle-resolved experiments on colloids \cite{konig2005,royall2008,mazoyer2011,ivlev,leocmach2012,tamborini2015,zhang2016,turci2016}. Further
experimental evidence of increasing numbers of LFS upon cooling is also found in metallic glassformers \cite{royall2015physrep,cheng2011}. Although not all these studies agree wholeheartedly on the role of structure, nor indeed upon the type of structure that is relevant, what is clear is that in many glassformers some structural change does occur.

In addition to this structural change, key to understanding the glass transition is the concept of dynamic heterogeneity, \emph{i.e.}~the existence of locally fast and slow regions in the system \cite{hurley1995,donati1998,ediger2000}. Determining the nature of these heterogeneities (probably) lies at the heart of understanding the glass transition since competing theoretical interpretations envisage different behaviour for dynamically heterogeneous regions. We may divide the process of dynamical arrest into two principle regimes: The \emph{known} and the \emph{unknown}. The known region corresponds to the first 3-5 decades of increase in structural relaxation time $\tau_\alpha$ relative to the normal liquid. This region is \emph{known} because particle resolved studies, i.e. molecular simulation and colloid experiment \cite{ivlev} allow us to determine the particle coordinates and precisely probe the behaviour of the system. In addition we have in this regime a theoretical framework, the so-called ``mode-coupling theory'' (MCT)~\cite{binder,goetze,charbonneau2005}, that is able to rationalize many aspects of the observed slowing down. Alas, this known regime ends, due to the relaxation timescales approaching the experimental or simulation timescales, just as things begin to get interesting from the point of view of understanding the glass transition. This dark zone starts around $T_\mathrm{mct}$, the critical temperature of MCT, at which the random first order transition (RFOT) theory \cite{lubchenko2007} predicts that the nature of the dynamical heterogeneities changes: For temperatures slightly above $T_\mathrm{mct}$ the so-called cooperatively rearranging regions (CRRs) that reflect the dynamical entities that are related to the regions of fast dynamics are relatively diffuse, whereas below $T_\mathrm{mct}$ they become more compact. For temperatures well below $T_\mathrm{mct}$ the size of the CRRs is predicted to grow until they reach the size of the system at the Kauzmann temperature $T_K$. One consequence of this scenario is that the configurational entropy becomes sub-extensive and a thermodynamic phase transition to an ideal glass
occurs is reached at $T_K$~\cite{lubchenko2007}.

Clear evidence in support of this scenario around the mode-coupling transition has been obtained from simulations that found that the $T-$dependence of the size of dynamically heterogeneous regions is non-monotonic for temperatures around $T_\mathrm{mct}$~\cite{kob2012}. This is thus consistent with the existence of more compact regions at deep supercooling. Although not all measures of dynamic lengthscales exhibit this temperature dependence \cite{flenner2012comment,kob2012reply,flenner2013}, and the feature might even be system-specific \cite{hocky2014pre}, a change in this dependence can be found for $T < T_\mathrm{mct}$ \cite{flenner2013}.  Some evidence for this behaviour has also been obtained in colloidal experiments \cite{nagamanasa2015,mishra2015}. However, since it is very hard to equilibrate colloidal systems suitable for particle-resolved studies at densities beyond the mode-coupling transition \cite{berthier2011,ivlev}, the use of this technique is somewhat limited. Of course simulations also have difficulties to access the interesting temperature regime but at present it appears possible to equilibrate at somewhat deeper supercooling with simulation than \emph{particle-resolved} colloid experiments \cite{ivlev,brambilla2009}. The very deep supercooling regime in which the CRR size are expected to grow significantly appears to elude even experiments on molecular liquids, \emph{i.e.}~systems for which one can reach timescales that are 14-15 decades larger than the relaxation time of the normal liquid \cite{ediger2000}. While for supercooled molecular liquids it is not possible to directly determine the size of the CRRs, a variety of indirect techniques have been employed, all of which indicate that at the experimental glass transition temperature $T_g$, at which the viscosity has reached $10^{12}$Pa$\cdot$s, the size of the CRRs is only around 10-100 molecules, thus a far cry from showing a divergence \cite{ediger2000,donth1982,cicerone1995,berthier2005,tatsumi2012,tracht1998,ashtekar2012,dalleferrier2007,yamamuro1998}.

There is also the standpoint that the slow dynamics is not related to an underlying thermodynamic transition but that the glassiness is due to ``dynamic facilitation''~\cite{chandler2010}. This approach predicts that so-called excitations (defect-like mobile regions) become increasingly sparser if temperature is decreased, yet remain essentially unaltered in their nature. For realistic systems the direct measurement of these excitations, 
which can be hard to accurately identify for off-lattice systems, is again limited to the $T\geq T_\mathrm{mct}$ regime, making it thus difficult to carry out stringent tests of this concept. But one can say that, like the agreement noted above for RFOT, facilitation also describes the dynamically heterogeneous regions in the regime accessible to simulation/colloid experiments rather well  \cite{keys2011}. Moreover both dynamic facilitation \cite{keys2013} and RFOT \cite{stevenson2006} have been shown to give reasonable predictions about the behaviour of molecular glassformers at supercoolings much greater than those accessible to simulation or colloid experiments. In short, at present we lack a way to definitively distinguish between competing theoretical approaches, even though the shapes of the dynamically heterogeneous regions --- becoming more compact and growing upon supercooling for RFOT, and staying similar in nature but becoming more sparse in the case of dynamic facilitation suggests that this might be a fruitful avenue \paddyspeaks{of research} to discriminate between these theories. The aim of this paper is to present a possible measurement method which may indeed enable such a discrimination.

Returning to the role of local structure, we observe that attempts to correlate it with dynamic heterogeneity, \emph{i.e.}, that the dynamically slow regions should be in the locally favoured structures, has met with some success \cite{dzugutov2002,coslovich2007,tanaka2010,sausset2010,malins2013jcp,hocky2014,jack2014}. However, the existence of such a correlation does by itself \emph{not} reveal a mechanism for arrest \cite{jack2014,charbonneau2013pre} and in any case is dependent on the system under consideration \cite{hocky2014}. A key limitation here is that direct detection of the LFS and dynamic heterogeneity is only possible in the first 3-5 decades of dynamic slowing accessible to colloids and computer simulation, i.e.~a region in the dynamic range where there might be cross-over effects from the high temperature to the low temperature dynamics.

In this contribution to the special issue ``Structure in glassy and jammed systems'', we present two related results.  Firstly, using simulations of a weakly polydisperse hard sphere system that is more deeply supercooled than those analysed to date \cite{royall2015,royall2014angell}. \paddyspeaks{This corresponds to a decade in the dynamics past the mode-coupling transition 
which we take as a packing fraction $\phi_\mathrm{mct}=0.58$ following van Megen \emph{et al.} \cite{vanmegen1998}).} We find an enhanced correlation between local structure and dynamically slow regions. This local structure, the icosahedron, is more compact than that observed previously \cite{royall2015}. Secondly, we introduce a new approach to characterise the shapes of cooperatively rearranging regions. In the regime accessible to our simulations, we find that their shapes are rather similar, and that the size of the CRRs seems to grow slower than the four-point dynamic correlation length often used, but find instead that it is rather well described by the lengthscale based on correlations in dynamic propensity \cite{dunleavy2015}.

\section{Lengthscales in glassy systems}

\subsection{Four-point  dynamic correlation length $\xi_4$} 

A variety of lengthscales have been proposed in connection with the glass transition. Some are based on structural quantities \cite{royall2015physrep} and others on dynamical heterogeneity \cite{karmakar2014}. Among the quantities underlying some dynamic lengthscales is the \emph{dynamic susceptibility} $\chi_4$ given by

\begin{equation}
\chi_4(t) = N \left[ \langle Q(t)^2\rangle -\langle Q(t) \rangle ^2 \right] \quad
\mbox{with} \quad Q(t):=\frac{1}{N} \sum_{n=1}^N w(|\textbf{r}_n(t)-\textbf{r}_n(0)|) 
\label{eqChi4}
\end{equation}

\noindent 
and which is obtained by integrating a four-point time-dependent density correlation function over volume \cite{lacevic2003}. Here $w(\textbf{r})$ is unity in the case that a particle has moved less than some distance (often, as is the case here, chosen to be $0.3\sigma$, where $\sigma$ is the diameter of the particles) and zero otherwise.  For a given temperature $T$ (or packing fraction $\phi$), the fluctuations (i.e., the susceptibility $\chi_4$) attain a maximum at a certain time $t=\tau_{h}$, and then die away, as shown in section~\ref{sectionResults}. The time $\tau_{h}$ has a value similar to \paddyspeaks{the structural relaxation time} $\tau_{\alpha}$. We note that the exact nature of the increase of the dynamic length scale is so far unclear, and in fact depends upon the system under consideration.

The quantity $\chi_4(t)$ is related to the collective cooperative dynamics of the particles. It can be easily generalized to the four-point dynamic structure factor $S_4(\textbf{k},t)$, a quantity that informs on this dynamics on a lengthscale defined by a wave-vector $\textbf{k}$, and that is  defined by

\begin{eqnarray*}
S_4(\textbf{k},t) &  =  & \frac{1}{N\rho} \langle \sum_{jl} \exp[i \textbf{k} \cdot \textbf{r}_j(0)]w(|\textbf{r}_j(0)-\textbf{r}_l(t)|)  \\
&\times & \sum_{mn} \exp[i \textbf{k} \cdot \textbf{r}_m(0)]w(|\textbf{r}_m(0)-\textbf{r}_n(t)|) \rangle,
\end{eqnarray*}

\begin{equation}
\label{eqS4}
\end{equation}

\noindent 
where $j$, $l$, $m$, $n$ are particle indices. For time $\tau_h$, the orientationally averaged version is $S_4(k,\tau_h)$. Its $k-$dependence may then be fitted with an Ornstein-Zernicke relation to obtain a dynamic correlation length $\xi_{4}$,

\begin{equation}
S_{4}(k,t_{{\rm h}})=\frac{S_{4}^{0}}{1+(k\xi_{4})^{2}},
\label{eqOZcrit}
\end{equation}

\noindent 
where $S_{4}^{0}$ is a fitting parameter. The obtained dynamic correlation length $\xi_{4}(T)$ can been fitted by a critical-like power-law with the mode-coupling temperature as the transition temperature~\cite{lacevic2003}. We emphasise, however, that a number of other fits are possible with a divergence at $T=0$~\cite{whitelam2004}, or at $T=T_0$ obtained from fitting the data with a Vogel-Fulcher expression~\cite{tanaka2010}, but also scalings which couple to the relaxation time can be done~\cite{flenner2013,flenner2009,flenner2011,kim2013}. As a consequence the divergence of $\xi_4$ can be associated with the temperature to where one imagines $\tau_\alpha$ to diverge --- $T_\mathrm{mct}, 0$ or $T_0$. Such a variation reflects the fact that in practice the observable range over which $\xi_4$ varies is less than one decade, so discussions of divergence require a considerable degree of faith. In other words, measuring $\xi_4$ from simulation data does (so far) not enable discrimination between different theories \cite{royall2015physrep}.

\subsection{Further dynamic lengthscales}

The results on the various systems that have been investigated so far indicate that the dynamic correlation length $\xi_4$ is almost always larger than the static correlation lengths from structural measures such as geometric motifs -- LFS -- or so called ``order-agnostic'' measures \cite{royall2015physrep}. This fact challenges the idea that the glass transition is a thermodynamic phenomenon: If there is some kind of underlying lengthscale associated with arrest, then structural lengthscales should surely be proportional to the dynamic lengthscales \cite{charbonneau2012,royall2015physrep,karmakar2014}.

To address this issue, one of us (CPR) has also introduced another length, based on information theoretic quantities. Details may be found in \cite{dunleavy2015}, but the essence is as follows: In the iso-configurational ensemble where many \paddyspeaks{(in our case 2048)} realisations are run independently from the same starting configuration \cite{widmercooper2006}, particles have a \emph{propensity} for being mobile - or immobile. As a consequence, across the ensemble of trajectories, particles have a probability
distribution of mobility. By measuring the spatial correlations in this mobility one can obtain a dynamic lengthscale $\xi_\mathrm{RG}$, where ``RG'' stands for radius of gyration of the so-obtained clusters. \paddyspeaks{Here clusters of particles were identified in terms of the mutual information between the mobility distribution of each particle across the isoconfigurational ensemble. In terms of the mutual information, it was found that the mobility of each particle is strongly correlated with a few neighbours, which defines the cluster. The number of neighbours was found to be maximised at a time very close to the structural relaxation time $\tau_\alpha$.} This dynamic lengthscale turns out to increase less than does $\xi_4$ and is in fact comparable to structural measures. Further details can be found in \cite{dunleavy2015}.

Another possibility to define a dynamic length scale has been proposed in Refs.~\cite{kob2012,hocky2014pre}. There the idea was to probe the details of the shape of the CRRs by looking at the relaxation dynamics of the system in the vicinity of an amorphous wall. Also in that case it was found that the extracted dynamic length increases at
a smaller rate than does $\xi_4$.

These two examples show that one can come up with definitions of dynamical length scales that are reasonable and that result in scales that have a $T-$dependence that is not too different from the one found for static scales. In the following we will thus follow this idea and see how the various scales compare with each other.

\section{Methods}
\label{sectionMethods}

Event-driven molecular dynamics (MD) simulations were carried out with the DynamO package \cite{bannerman2011}. We used an equimolar five component mixture of hard spheres whose polydispersity is 8\%. \paddyspeaks{Each particle has the same mass $m$. The diameters of each species are then $[0.888\sigma, 0.95733\sigma, \sigma, 1.04267\sigma, 1.112\sigma]$. In the following we take $\sigma$ as the unit of length. For the unit of time, we take $\sqrt{m \sigma / k_B T}$.}

We equilibrated for at least $30$ $\tau_{\alpha}$ and sampled for at least a further $30$ $\tau_{\alpha}$, where $\tau_{\alpha}$ is the structural relaxation time which we determined by fitting a stretched exponential \wkspeaks{to the (collective) intermediate scattering function at the main peak of the static structure factor.} Two system sizes were considered, $N=1372$ and $N=10976$, \wkspeaks{but no relevant differences have been found between their relaxation dynamics. }\paddyspeaks{Further details may be found in \cite{royall2015}.}

The packing fractions we have considered are $\phi=0.19$, 0.42, 0.50, 0.54, 0.56, 0.57, 0.575, 0.58, and 0.585. 
\paddyspeaks{Each state point is prepared independently and we analyse typically 3000-10,000 configurations.}
Following Ref.~\cite{royall2015} we will present our data as a function of the compressibility factor $Z$ instead of the packing fraction $\phi$. For this quantity we take the Carnahan-Starling form $Z_\mathrm{cs}$, noting that for our system, where we can obtain data of sufficient quality, we see no deviation from the Carnahan-Starling expression. Thus $Z_\mathrm{cs}$ is given by

\begin{equation}
Z_\mathrm{cs}=\frac{1+\phi+\phi^2-\phi^3}{(1-\phi)^3} \quad .
\end{equation}

\noindent 
We identify local structure using the topological cluster classification (TCC) \cite{malins2013tcc} and focus on three particular structures which we believe are important in this system \cite{dunleavy2015} and that are described in more detail in Sec.~\ref{sectionLocalStructure}.

\wkspeaks{In the following we will refer to ``fast'' and ``slow'' particles. The label ``fast'' is given to 
particles which move more than $0.3\sigma$ as measured from their squared displacement on the time scale $\tau_h$, which corresponds to around 50\% of the particles. All other particles are defined as ``slow''.}

\section{Results}
\label{sectionResults}

\subsection{Global dynamics}
\label{sectionGlobalDynamics}

\begin{figure}
\centering
\includegraphics[width=\textwidth]{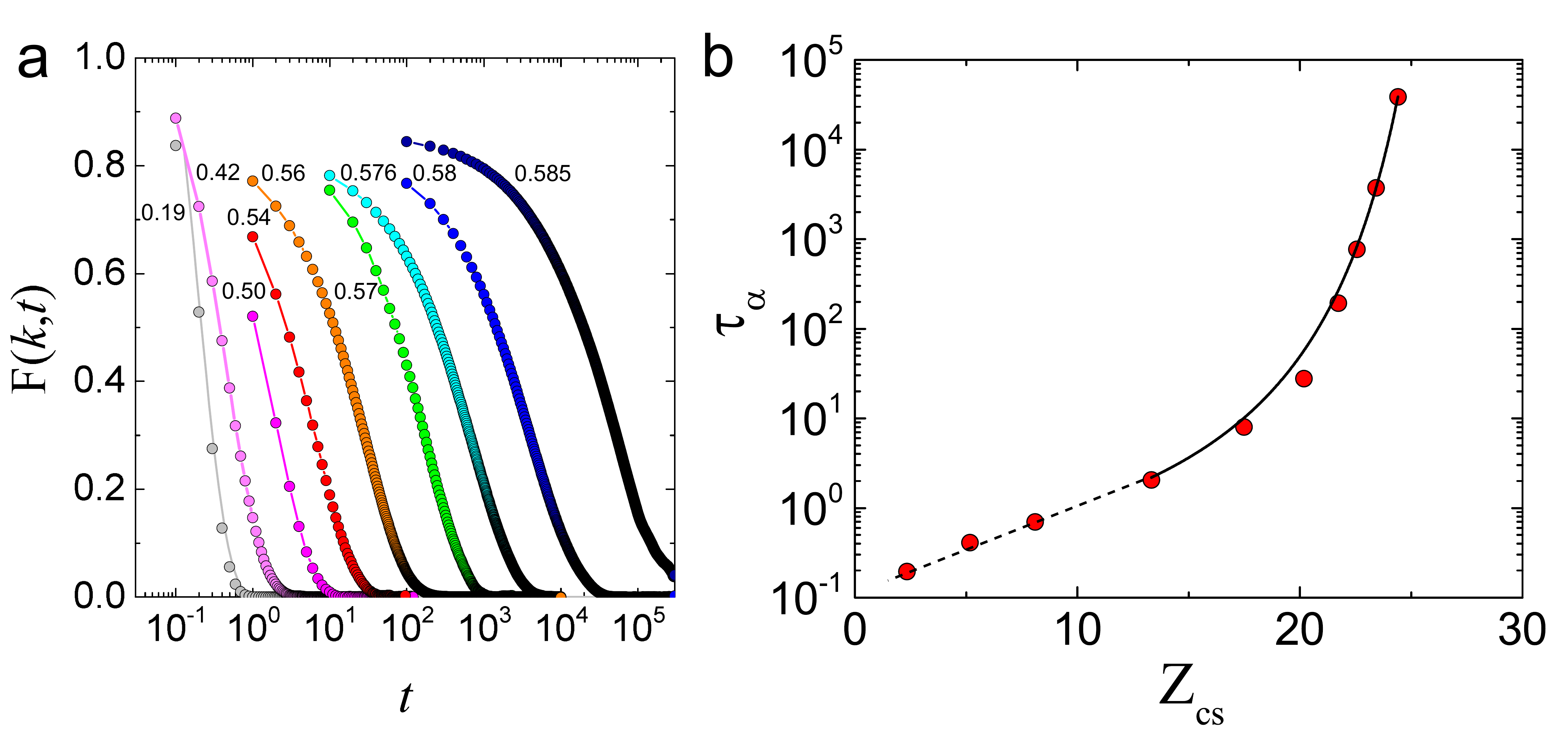}
\caption{
(Colour online) Relaxation dynamics of the weakly polydisperse hard sphere system. 
(a) Intermediate scattering function $F(k,t)$.  The
wavevector $k$ is taken close to the main peak in the static structure factor. The volume fractions of the state points shown are indicated in the figure.
(b) Relaxation times $\tau_\alpha$ as a function of $Z_\mathrm{cs}$. The solid line is a VFT fit with $Z_\mathrm{cs}$ as the control parameter [Eq. \ref{eqVFT}]. 
}
\label{figDynamics}
\end{figure}

In Fig.~\ref{figDynamics}(a) we show the time dependence of the intermediate scattering function for different values of $Z_\mathrm{cs}$. By fitting the $\alpha-$relaxation part of these curves with a Kohlrausch-Williams-Watts form we have determined the $\alpha-$relaxation time $\tau_\alpha$ which is shown in Fig.~\ref{figDynamics}(b). The $Z_\mathrm{cs}-$dependence of $\tau_\alpha$ can be fitted with a Vogel-Fulcher-Tammann-like expression of the form

\begin{equation}
\tau_{\alpha}=\tau_{0}\exp\left[ \frac{\tilde{E}_{{\rm a}}}{k_{{\rm B}}(Z_0-Z)} \right]
\label{eqVFT}
\end{equation}

\noindent 
where $Z_{0}~(\simeq Z_{{\rm K}}$) and $\tilde{E}_{{\rm a}}$ is a measure of the fragility. For our data, we find $Z_0=28.73$, corresponding to a packing fraction $\phi=0.609$ with a fragility parameter $\tilde{E}_\mathrm{a}/k_\mathrm{B}=69.74$.

\begin{figure}
\centering
\includegraphics[width=\textwidth]{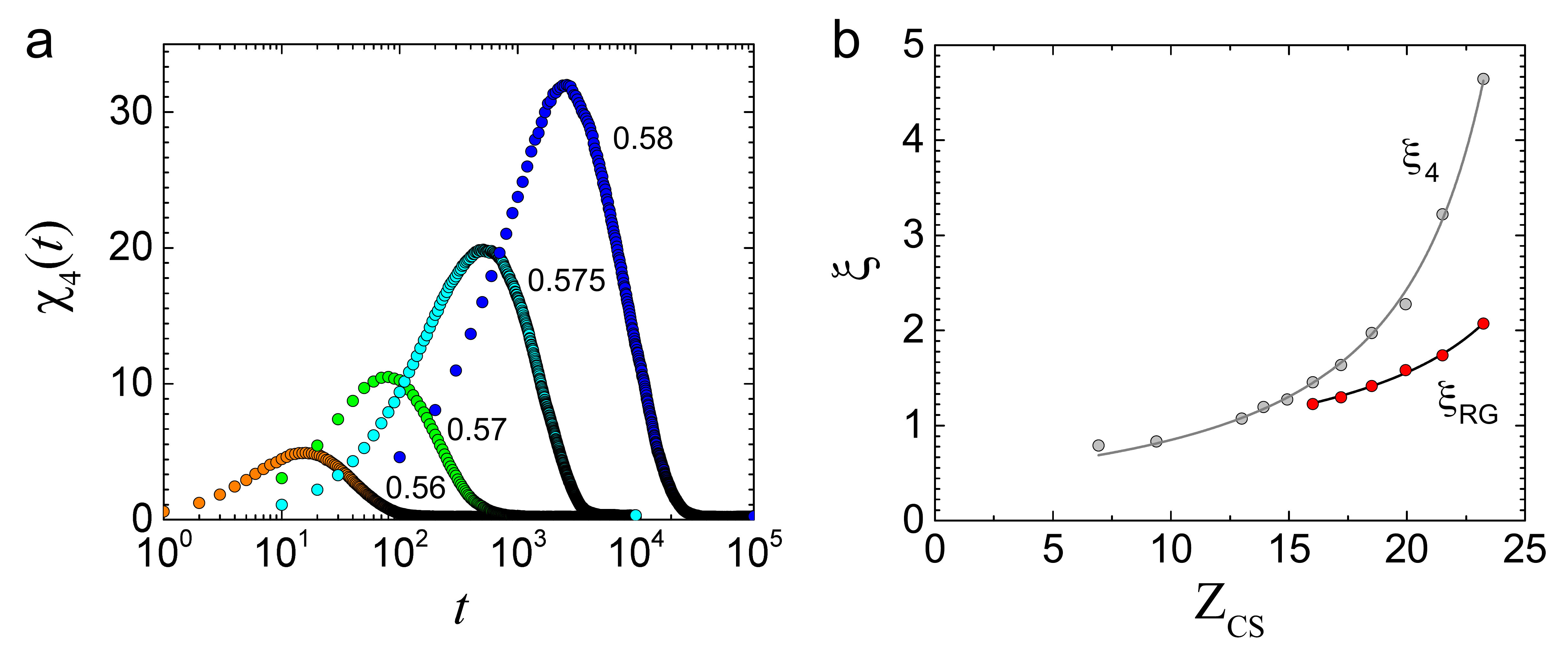}
\caption{
(Colour online) $(a) \chi_4(t)$ for increasing $Z_\mathrm{cs}$ or $\phi$. Volume fractions are indicated. (b) Dynamic
lengthscales fitted with Eq.~(\ref{eqChiara}) with parameters appropriate to the four-point correlation length $\xi_4$ and the mutual information based length $\xi_\mathrm{RG}$.
\paddyspeaks{Data points in (b) are taken from \cite{dunleavy2015}.}
}
\label{figDynHet}
\end{figure}

We now consider dynamic heterogeneity. In Fig.~\ref{figDynHet}(a) we plot $\chi_4(t)$ as defined by Eq.~(\ref{eqChi4}). This quantity exhibits the characteristic increase upon approaching the glass transition. In Fig.~\ref{figDynHet}(b) we plot the length obtained for $\xi_4$ and $\xi_\mathrm{RG}$ as obtained in Ref.~\cite{dunleavy2015}.  These two lengthscales can be fitted with the function

\begin{equation}
\xi(Z)=\xi_0 \left( \frac{1}{Z_0-Z} \right)^\frac{1}{d-\theta}
\label{eqChiara}
\end{equation}

\noindent 
which is inspired by RFOT \cite{cammarota2009} and is here mainly chosen for simplicity. Here $d$ is dimension and $\xi_0=48.8$ and $5.96$ and $\theta=2.28$, and $1.38$ for $\xi_4$ and $\xi_\mathrm{RG}$ respectively. This provides us with estimates for the lengthscales at arbitrary $\phi$. We emphasise that Eq. (\ref{eqChiara}) should be interpreted merely as a fit, i.e.~we do not wish to imply a physical interpretation here.

\begin{figure}
\centering
\includegraphics[width=8cm]{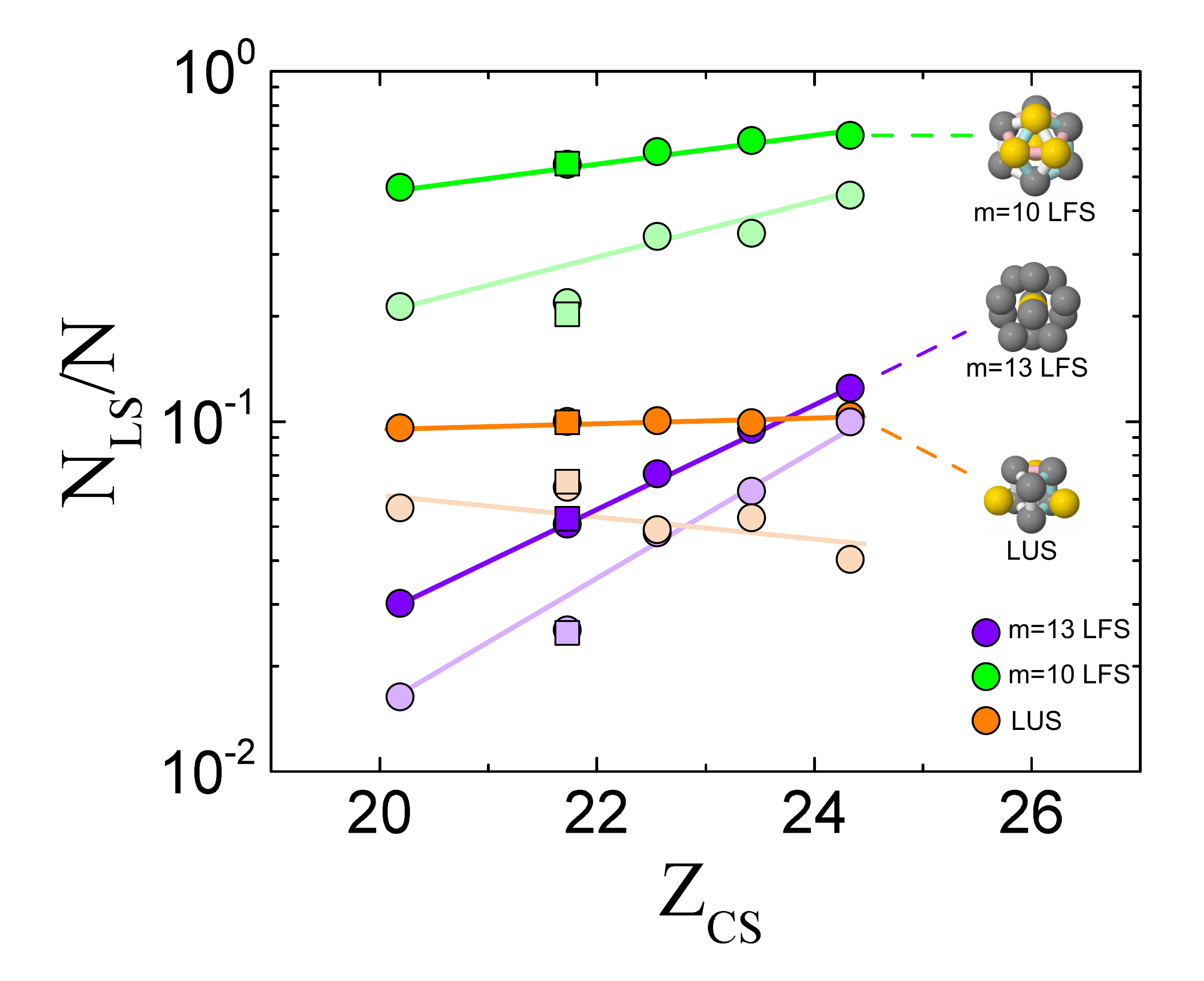}
\caption{(Colour online) 
Fraction of particles in locally favoured --- and unfavoured --- structures as a function of the compressibility factor $Z_\mathrm{CS}$. Purple: icosahedra (m=13)  (LFS); green: defective icosahedra (m=10) (LFS); orange: triangular prisms (LUS). 
\paddyspeaks{Pale} symbols correspond to particles in LFS and LUS which are slow and fast, respectively.  Round symbols are N=1372, square are N=10976. Solid lines are to guide the eye. \paddyspeaks{Error bars due to statistical differences between runs are smaller than the symbol size.}
}
\label{figNLS}
\end{figure}

\subsection{Local structure: favoured and unfavoured}
\label{sectionLocalStructure}

Here we probe how certain particular geometric structures depend on the packing fraction since this information is useful to understand the nature of the local structures.

In previous work, a ten-membered 
structure has been identified as being long-lived and rather prevalent in this system \cite{royall2015}. This locally favoured structure is a defective icosahedron (an icosahedron missing three particles)~\cite{malins2013tcc}. In Fig.~\ref{figNLS} we show the $Z_{\rm CS}-$dependence of these defective icosahedra as well as the population of full 13-membered ones. Furthermore the system contains also local structures that are {\it unstable}, i.e.~these \emph{locally unfavoured structures} (LUS) are dynamically fast. One of us has found that these structures take the form of a triangular prism with some additional particles, shown as a cartoon in Fig. \ref{figNLS}, and the $Z_{\rm CS}-$dependence of its concentration is included in the figure as well. It is believed that the four-membered rings from which such structures are comprised are particularly mechanically unstable \cite{dunleavy2015}.

We thus consider three species of local structures which we count as follows. The nine-membered LUS are simply expressed as the fraction of particles in an LUS. To define the concentration of the LFS (full or defective icosahedra) we consider all the particles that are in one of these structures. Note that since a particle can belong to more that one of these structures, the total concentrations of structures can add to more than 1.0. While there are other possibilities for counting the local structures, we emphasise that these have no effect on our conclusions, and obvious choices have very little quantitative impact,

In Fig.~\ref{figNLS} we show the fraction of particles belonging to the different local environments as a function of the compressibility factor $Z_\mathrm{CS}$, accessing now values of $Z_\mathrm{CS}$ that so far have been unattainable. The filled symbols are for the total populations whereas the open symbols are the fractions of particles that are in LFS and move slowly or in LUS and fast. [Note that according our criteria (see section \ref{sectionMethods}), about half of the particles are slow]. From this figure we can draw the following conclusions: Firstly, the number of full icosahedra (\paddyspeaks{purple} filled symbols) increases strongly with increasing $Z_\mathrm{CS}$. Secondly, the fraction of particles that are both slow and in these full icosahedra (blue open symbols) increases with $Z_\mathrm{CS}$ and account at the highest densities for about 80\% of the particles in the icosahedra, i.e.~well beyond the trivial statistical expectation of 50\%. From this we thus can conclude that with increasing glassiness of the system an increasing fraction of the slow particles \paddyspeaks{forms} part of a full icosahedron. Thirdly, the number of LUS (orange filled symbols) is basically independent of $Z_{CS}$ and the number of particles that are in the LUS {\it and} are fast (orange open symbols) makes up only about 50\% of the total and show a weak trend to {\it decrease}. This indicates that there seems to be rather little correlation between LUS and dynamically fast particles and thus the life time of the LUS is not short because its constituent particles are fast.

\paddyspeaks{We emphasize that we do not expect the trend of increasing populations of icosahedra to continue all the way to random close packing. Provided the system can be equilibrated, one of us (CPR) has shown that there is an increase in fivefold symmetry, but that when it falls out of equilibrium, the degree of fivefold symmetry is much reduced; and tetrahedra symmetry is found instead \cite{royall2014arxiv}. Thus when the system falls out of equlibrium (\emph{i.e}. well before random close packing), we expect a reduction in the quantity of icosahedra. Even in equilibrium, it may saturate at some intermediate equilibrium state point, which could perhaps be related to a fragile-to-strong transition.}

\begin{figure}
\centering
\includegraphics[width=\textwidth]{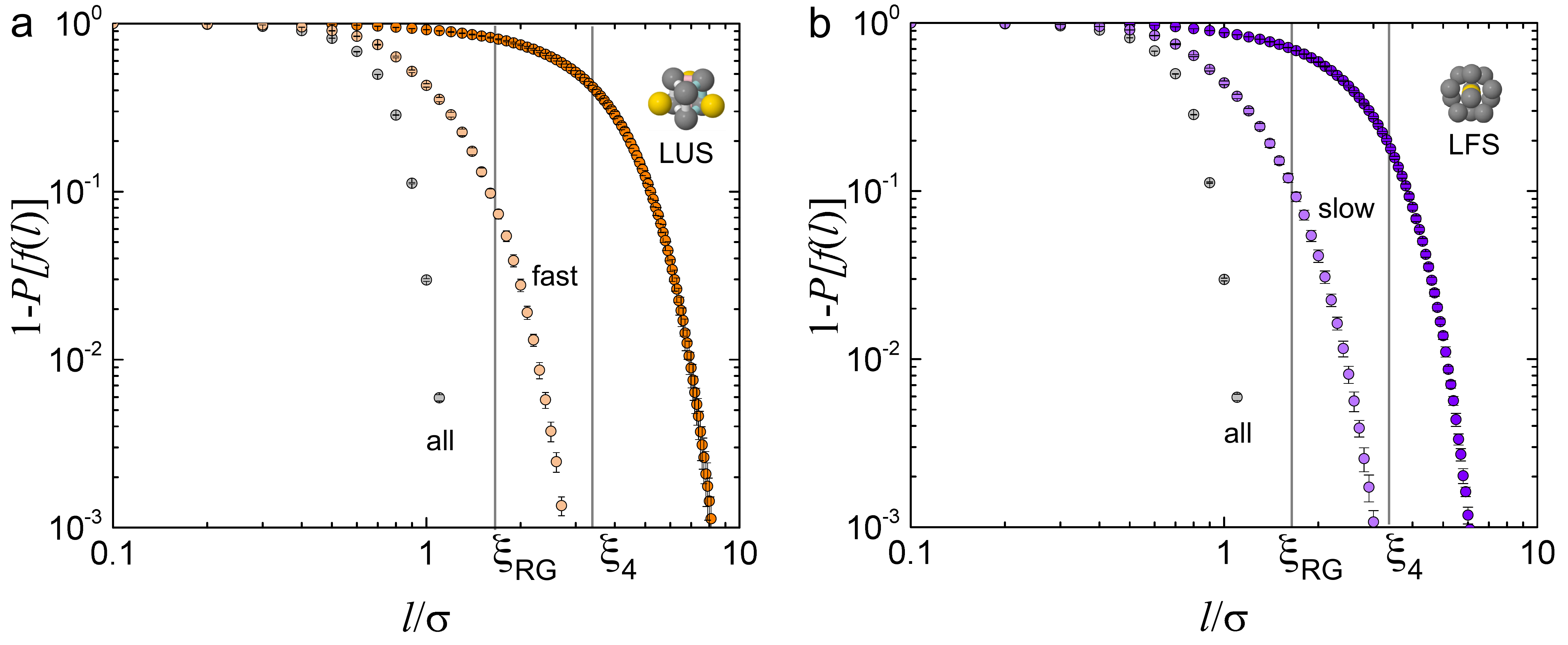}
\caption{(Colour online) 
Probability that a box of size $l$ contains a particle with a certain ``property'' (see main text for details).  Packing fraction $\phi=0.576$. The vertical lines show the lengthscales $\xi_4$ and $\xi_\mathrm{RG}$ from Ref.~\cite{dunleavy2015}.  
(a) All particles (grey), fast particle (light orange), and locally unfavoured structures (orange).  
(b) All particles (grey), slow particles (light purple), and locally favoured structures (purple). 
\paddyspeaks{Error bars correspond to the standard deviation across eight independent simulations.}
}
\label{figNN}
\end{figure}

\subsection{Shapes of dynamically heterogeneous regions}

We now introduce a new approach to determine the shape of dynamically heterogeneous regions and from this a new dynamic length scale. This method is motivated by the fact that in Ref.~\cite{kob2012} evidence was found that the shape of the CRRs depends on the packing fraction (or temperature), in qualitative agreement with the prediction by RFOT~\cite{stevenson2006}.  Since these structures might have fractal-like properties it is not very appropriate to try to characterize their size with a naively chosen length scale and hence we propose here 
\paddyspeaks{instead an approach }
inspired from the analysis of fractal-like objects. In particular we analyse their structure using a box counting algorithm. For this we determine $f(l)$, the probability that a cubic box of size $l$ that is placed randomly in the sample contains at least one particle that has a certain property (to be defined below). Since the values of $f(l)$ that are relevant for the present study are close to unity, it is useful to discuss not $f(l)$ but $1-f(l)$, i.e.~the probability that a box does {\it not} contain any particle with the selective property. Thus in contrast to the various length scales that can be found in the literature and that focus on the structures of interest (LFS, LUS, ...), we look here at the {\it complementary} space of these structures. \paddyspeaks{Before presenting the results, we discuss briefly the interpretation of the method we use here. In short, we measure the ``fuzziness'' of the various regions. This is characterised by the rapidity with which $1-f(l)$ falls as $l$ is increased. That is to say, for compact, $d=3$ regions with well-defined (spherical) boundaries, we expect that $1-f(l)$ falls precipitously. The slower $1-f(l)$ falls, the more diffuse the region.}

In Fig.~\ref{figNN}, we plot $1-f(l)$ for a variety of choices for the property of the particles used to define $f(l)$. The data is for a relatively elevated packing fraction, $\phi=0.575<\phi_\mathrm{mct}\approx 0.58$, but below we will discuss how these results depend on $\phi$.  The grey data points correspond to the box-counting for all particles. As expected for dense assemblies of spheres, once the box size becomes comparable to the particle size, the probability that a box is empty plummets. In the case of fast  particles, Fig.~\ref{figNN}(a), somewhat larger boxes can be inserted, but when the box size $l$ is around that of the mutual information length $\xi_\mathrm{RG}$, the chance to insert an empty box falls quite quickly. This decrease is, however, not as fast as in the case of all particles which indicates that the complementary space of
the set of fast particles has an irregular structure, a result that does not surprise given the irregular shape of the fast regions. However, since there is no indication for a power-law in our data, we can conclude that \emph{at the lengthscales we access,} this complimentary space is not fractal-like. Once $l$ has reached the size of $\xi_4$, the probability has basically gone to zero, showing that the typical distance between clusters of fast particles is significantly smaller than $\xi_4$. Also included in the graph is the data for the LUS. We see that this probability starts to decay to zero in a manner that is very similar to the one we observed for the fast particles. The main difference is that the length scale at which this decay starts is now a factor 2-3 larger, which is reasonable since these structures account only for about 10\% of the particles whereas the fast particles account for 50\% (see Fig.~\ref{figNLS}).

In Fig.~\ref{figNN}(b) we show the corresponding probabilities for the slowly moving particles and the LFS. Note that, since about 50\% of the particles are slow and 50\% are fast, the regions that are fast are basically the complement of the regions that are slow, and {\it vice versa}. We recognize that the $l-$dependence of the curve for the slowly moving particles is quite similar to the one for the LFS and that the latter decays at a length scale that is about a factor of four larger than the one for the fast particles. In view of the relative concentrations of these structures/particles (see Fig.~\ref{figNLS}) this factor is again quite reasonable. What is somewhat surprising is the curves are remarkably similar to the ones of the fast particles/the LUS. This tells us that, under our criterion for fast/slow particles (see section \ref{sectionMethods}), the CRRs, \emph{i.e.} the shapes of fast and slow regions are not very different --- and somewhat irregular.

\begin{figure}
\centering
\includegraphics[width=\textwidth]{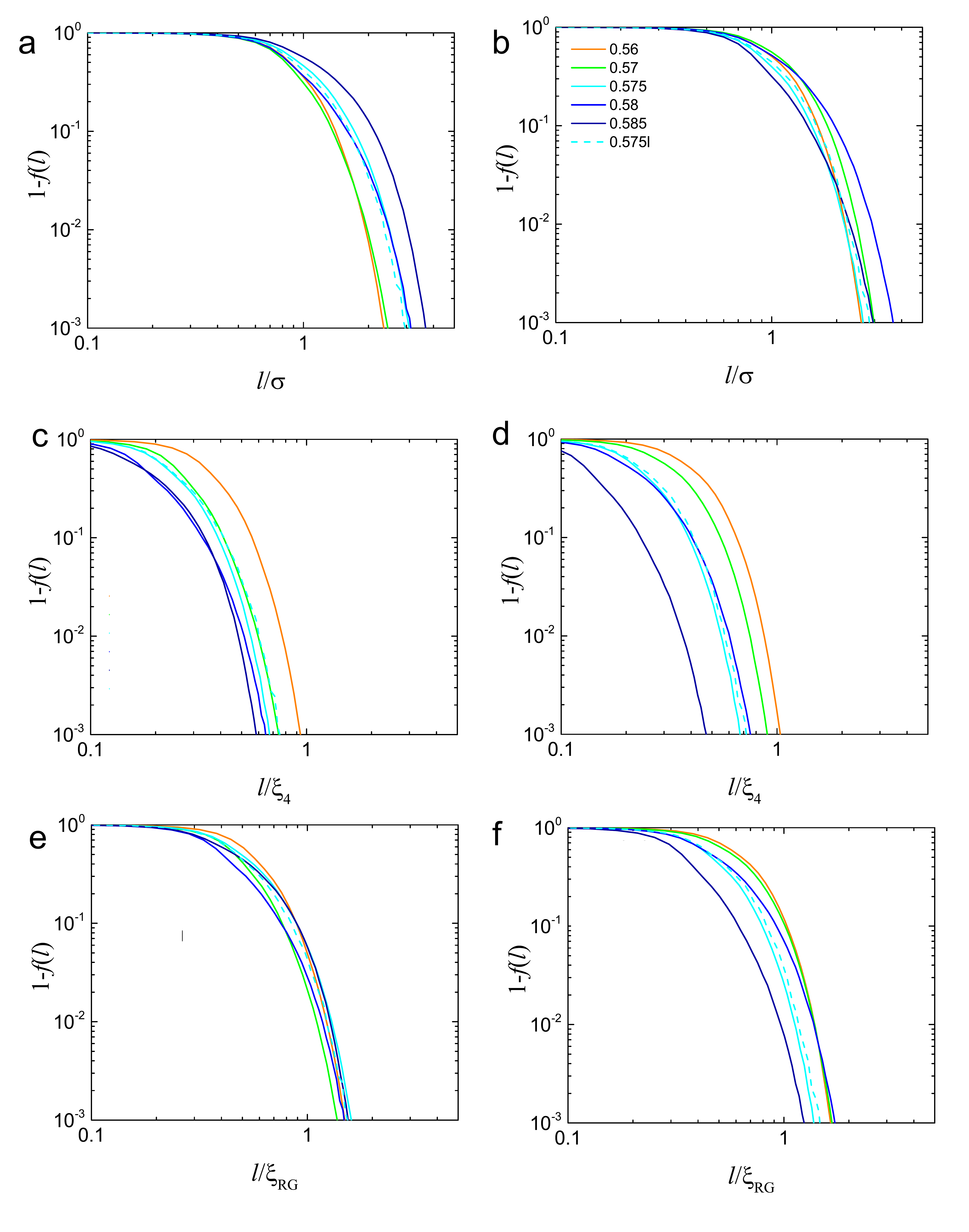}
\caption{(Colour online) 
Probability that a box of size $l$ does not contain a fast/slow particle, left/right column, respectively, for different packing fractions $\phi$. The solid lines are for $N=1372$ particles and the dashed line for $N=10976$ \paddyspeaks{($\phi=0.575$), which corresponds to a simulation box of approximately double the side of the smaller systems.} In panels (a) and (b) $l$ is measured in units of $\sigma$, in (c) and (d) in units of the four-point dynamic correlation length $\xi_4$, and in (e) and (f) in units of the propensity correlation length $\xi_\mathrm{RG}$.  Packing fractions are indicated in (b) and hold for all panels.  
}
\label{figCRRshapesPhi}
\end{figure}

Finally, we investigate in Fig.~\ref{figCRRshapesPhi} how the shape and the size of fast and slow regions changes with supercooling. The left and right column of the figure is for the fast and slow particles, respectively. The full lines are for $N=1372$ and the dashed line for $N=10976$. The fact that for $\phi=0.585$ the curves for these two sizes are very close together shows that these are no significant finite size effects and that the accuracy of the data is quite high. Note that we present here data in a $\phi-$range that enclosed the MCT-transition and in which the relaxation times change by more than four decades.

In panels (a) and (b) of Fig. \ref{figCRRshapesPhi} we plot the probability to find an empty box as a function of its size (in units of $\sigma$). We recognize that this
distribution depends only weakly on the packing fraction, which is in agreement with the usual finding that the length scales associated with fast/slow domains do not change strongly with the degree of supercooling. Panel (a) shows that with increasing packing fraction the curves have the tendency to move to the right, indicating that the shape of the region with the fast moving particles become better defined and take up less space, also this in qualitative agreement with previous findings.  The data from panel (b) tells a similar message, namely that the shape of the curves, and hence the shape of the domains, changes only weakly with $\phi$. There is, however, one particularity: The sequence of the curves is non-monotonic in the packing fraction in that the curves first move to the right, as expected, before jumping back again to the left (see data for $\phi=0.575$). This behavior, if real, is evidence that the size of the fast regions first increases and then shrinks again, in qualitative agreement with the results from Ref.~\cite{kob2012}.

Panels (a) and (b) report the length scale in units of $\sigma$. The other panels in Fig.~\ref{figCRRshapesPhi} show the same distribution using two dynamic lengthscales as a yardstick. Since these lengthscales depend on $\phi$ these graphs should allow us to see whether or not these dynamic length scales are able to probe the size of the fast/slow domains. Upon rescaling the box length by the four-point dynamic correlation lengths, $\xi_4$ in Fig.~\ref{figCRRshapesPhi}(c,d) and dynamical propensity based length $\xi_\mathrm{RG}$, Figs.~\ref{figCRRshapesPhi}(e,f), we find that the latter achieves a better data collapse. The four-point dynamic correlation length $\xi_4$ increases faster than any lengthscale associated with $1-f(l)$. This is revealed by the fact that the  $1-f(l)$ curves moving to \emph{smaller} values of $l/\xi_4$ upon deeper supercooling. Remarkably, by scaling $l$ by $\xi_\mathrm{RG}$, we find that $1-f(l/\xi_\mathrm{RG})$ changes little. It is thus possible to interpret $\xi_\mathrm{RG}$ as an appropriate lengthscale when considering CRRs. While we emphasise that ``one swallow doesn't make a spring'', we observe that in Fig. \ref{figCRRshapesPhi}(e), it is possible to convince oneself that perhaps the fall-off of $1-f(l)$ is less steep at deeper supercooling. This would imply more diffuse CRRs as predicted by RFOT \cite{lubchenko2007}.

\section{Conclusions}

We have carried out a structural analysis of configurations of a weakly polydisperse hard sphere system at a deeper supercooling than previously accessed, a decade in dynamics past the mode-coupling transition. We  reveal the emergence of a locally favoured structure, whose population is small at weaker supercooling. This is the ``canonical'' icosahedron predicted long ago \cite{frank1952}. At these deeper supercoolings, the LFS correlate well with the slow particles. This tells us that, under our criterion for fast and slow particles, there is effectively a change in LFS from defective icosahedra at weaker supercooling \cite{royall2015} to full icosahedra, indicating that these LFS are a feature of deeper supercooling than that at which most of the simulation (and colloid experiment) investigations have been made. We note, however, that the present
results might be restricted to hard-sphere like systems that are (basically) mono-disperse, \emph{i.e.}~liquids in which there is no finite attraction/repulsion or chemical order that could give rise to LFS that have a more complex shape/composition.

We have developed a new means to measure the shapes of CRRs which is inspired by the analysis of fractal objects.  In the regime of supercooling accessible to the present simulations, the CRR shape, related to the manner $1-f(l)$ decays, appears to change little. The absolute size of these regions increases, as usual, only moderately with increasing packing fraction. When these distributions are plotted as a function of a dynamic lengthscale based on mutual information, $\xi_\mathrm{RG}$ \cite{dunleavy2015}, we find a \paddyspeaks{reasonable} master curve which indicates that the shape of the CRRs does not depend strongly on $\phi$. This master curve also shows that $\xi_\mathrm{RG}$ does indeed capture the size of the CRRs, which is not the case for $\xi_4$ since the latter does not lead to a data collapse. Instead our analysis indicates that with supercooling $\xi_4$ increases rather faster than the size of the CRRs.

One of the motivations for this work was to see whether it is possible to detect in a bulk system any sign for a non-monotonic dynamic length scale.  Our analysis gives no clear signature for such an effect, although certain quantities, e.g.~the distribution for the slow particles shown in Fig.~\ref{figCRRshapesPhi}(b), might hint that such a non-monotonic behavior does indeed exist. {\it If}~this is indeed the case this would thus open the door for analysing in a simple manner other glass-forming systems {\it in the bulk}, thus avoiding the somewhat complex setup used in Ref.~\cite{kob2012} (freezing in of an amorphous wall).

The possible absence of the non-monotonic behavior in the dynamical length scale for the present system might not be a total surprise since it has been found that a binary Lennard-Jones model did not show such a non-monotonic behavior either~\cite{hocky2014}.  Of course, it is entirely possible that, at deeper supercoolings, we may find in the present system some non-monotonicity in the correlation length. On the other hand it should be recalled that this feature might be closely related to the fragility of the glass-former (in the sense of Angell) since fragile systems show a more pronounced change in the activation energy, \emph{i.e.}~a change of slope in an Arrhenius plot of the relaxation
time. This change of slope is one of the main features of MCT (and therefore of RFOT) that predicts in effect a singularity in the dynamics, which, however, is cut-off by hopping processes that are not taken into account in the simplest forms of the theory. Thus RFOT can be expected to work better for fragile systems and in view of the fact that the harmonic sphere system of Ref.~\cite{kob2012} is more fragile than the polydisperse hard sphere system considered here, which in turn is more fragile than the binary Lennard-Jones system of Ref.~\cite{hocky2014}, the absence of a clear signature for a non-monotonic behavior is not that surprising, in agreement with other work~\cite{berthier2012pre}. To summarize we can conclude that the approach presented here allows to gain information on the properties of the CRRs without making
any assumptions on their nature or shape. Therefore it will be very interesting to see what this method finds for other glass-forming systems notably for those for which one seeks to understand well the nature of the relaxation dynamics and the associated length scales.

\ack
Andrew Dunleavy and Karoline Wiesner are greatly thanked for earlier work which inspired some of the results presented here.  CPR acknowledges the Royal Society, and Kyoto University SPIRITS fund. European Research Council (ERC consolidator grant NANOPRS, project number 617266) for financial support. W. Kob acknowledges support from the Institut Universitaire de France.

\section*{References}


\end{document}